\documentclass[twocolumn,trackchanges]{aastex631}

\hypersetup{linkcolor=blue,citecolor=blue,filecolor=brown,urlcolor=blue}

\newcommand{\kms}{\,km\,s$^{-1}\,$}

\begin{document}

\title{Confirmation of an anomalously
low dark matter
content for the galaxy NGC1052-DF4 from
deep, high resolution continuum spectroscopy}



\correspondingauthor{Zili Shen}
\email{zili.shen@yale.edu}

\author[0000-0002-5120-1684]{Zili Shen}
\affiliation{Department of Astronomy, Yale University, 
New Haven, CT 06511, USA}

\author[0000-0002-8282-9888]{Pieter van Dokkum}
\affiliation{Department of Astronomy, Yale University, New Haven, CT 06520, USA}

\author[0000-0002-1841-2252]{Shany Danieli}
\altaffiliation{NASA Hubble Fellow}
\affiliation{Department of Astrophysical Sciences, 4 Ivy Lane, Princeton University, Princeton, NJ 08544, USA}

\begin{abstract}
NGC1052-DF4 was found to be the second ``galaxy lacking dark matter'' in the NGC1052
group, based on its velocity dispersion of $\sigma_{\rm gc}=4.2^{+4.4}_{-2.2}$\,\kms\
as measured from the radial velocities
of seven of its globular clusters. Here we verify this result by measuring the stellar
velocity dispersion of the galaxy.
We observed the diffuse stellar light in NGC1052-DF4 with the Keck Cosmic Web Imager (KCWI)
in its highest resolution mode, with $\sigma_{\mathrm{instr}}\approx 7$\kms.
With a total science\,+\,sky exposure time of 34\,hrs, the resulting spectrum is exceptional both in its spectral resolution and its S/N ratio
of 23\,\AA$^{-1}$.  We find a stellar velocity dispersion
of $\sigma_{\rm stars} = 8.0^{+2.3}_{-1.9}$\,\kms,
consistent with the previous measurement from
the globular clusters. Combining both
measurements gives a fiducial dispersion of
$\sigma_{\rm f} = 6.3_{-1.6}^{+2.5}$\,\kms.
The implied dynamical mass within the half-light radius is $8_{-4}^{+6}  \times 10^7 M_{\odot}$. The expected velocity dispersion of NGC1052-DF4 from the stellar mass alone is $7 \pm 1$\kms, and
for an NFW halo that follows the stellar mass -- halo mass relation and the halo mass -- concentration relation, the expectation is $\sim
30$\,\kms. 
The low velocity dispersion rules out a normal NFW dark matter halo, and we confirm
that NGC1052-DF4 is one of at least two galaxies
in the NGC1052 group that have an anomalously low
dark matter content.
While any viable model for their
formation should explain the
properties of both galaxies, we note
that NGC1052-DF4 now poses the largest challenge as it has the most stringent constraints on its dynamical mass.
\end{abstract}



\section{Introduction} \label{sec:intro}
\begin{figure*}[htbp!]
\plotone{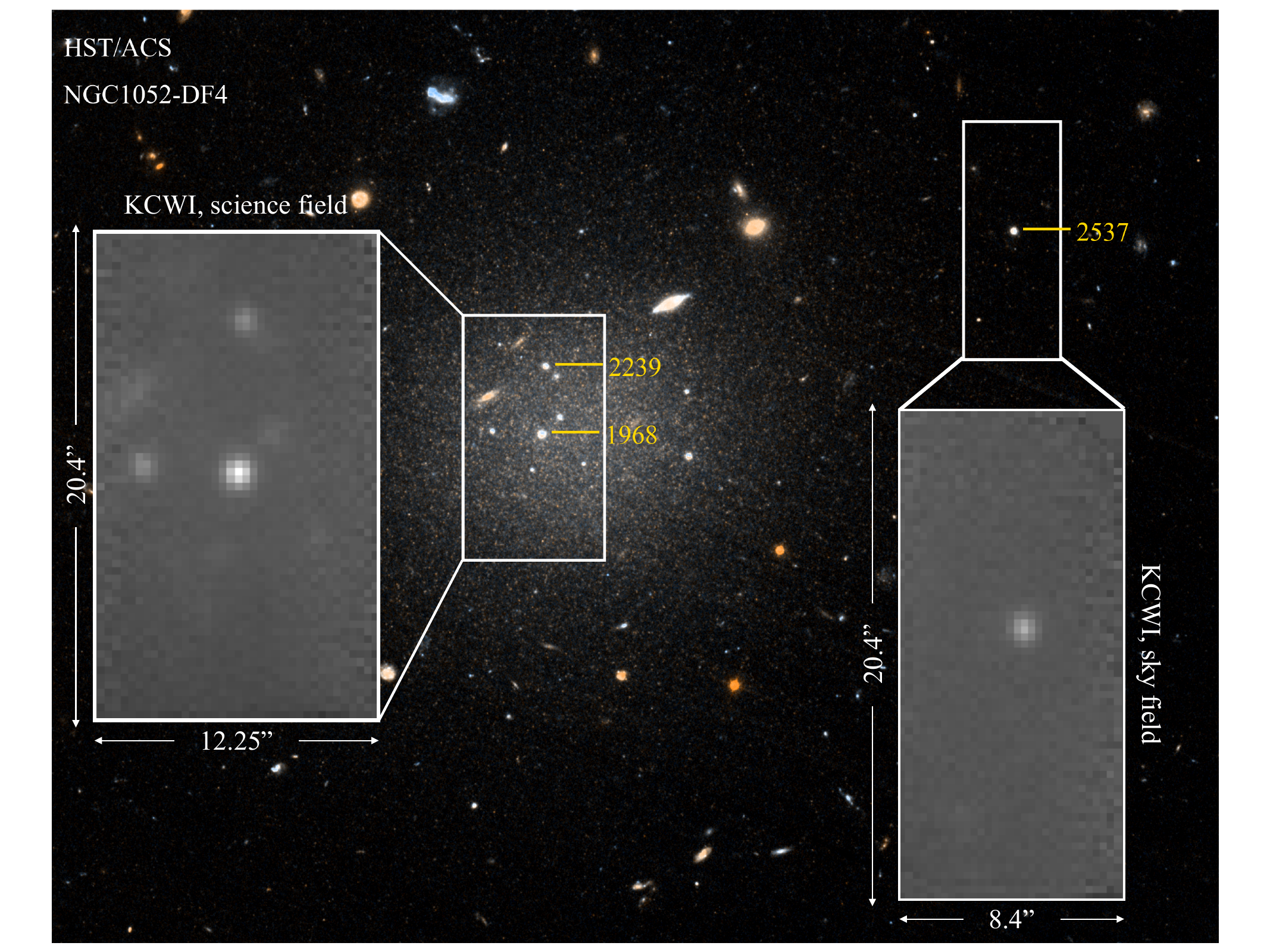}
\caption{\textit{HST/ACS} color image of NGC1052-DF4, created from $V_{\mathrm{606}}$ and $I_{\mathrm{814}}$ bands. The white rectangles represent the KCWI pointings: due to dithering, the science field that covers the diffuse light of the galaxy is bigger than the sky field, used for modeling the sky background. The inset panels show the KCWI data cubes flattened in the spectral dimension. The mean radius of the science field covers 95\% of $R_e$.
\label{fig:FOV}}
\end{figure*}

NGC1052-DF4 (or DF4) was the ``second galaxy missing dark matter'', with an extremely low velocity dispersion that is consistent with its stellar mass alone \citep{vanDokkum2019ApJL}.
Its velocity dispersion of $4.2_{-2.2}^{+4.4}$\kms was measured from seven globular clusters that are associated with DF4.
It shares many puzzling properties with the ``first galaxy missing dark matter'', NGC1052-DF2 \citep[DF2,][]{vanDokkum2018Natur}.
Located in the NGC1052 group ($D_{\mathrm{DF2}}=21.7$ Mpc, Appendix \ref{appdx}; $D_{\mathrm{DF4}}=20.0$ Mpc, \citealt{Danieli2020ApJL}), both galaxies have low velocity dispersions that imply an extreme deficit of dark matter, and host a population of overluminous globular clusters \citep{Shen2021ApJ}.
In addition, DF2 and DF4 have the same color in their globular clusters and the diffuse stellar light \citep{vanDokkum2022ApJL}.

Many scenarios have been proposed to explain why a single galaxy might appear to be lacking dark matter in the present day.
One {scenario} is that DF2 is on a special orbit, such that dark matter was tidally stripped by close encounters with NGC1052 \citep{Ogiya2018MNRAS,Nusser2020ApJ,Maccio2021MNRAS,Jackson2021MNRAS,Ogiya2022MNRAS,Moreno2022NatAs}. Another idea is that the galaxy is not pressure supported but is rotating, and viewed nearly face-on. \citet{Lewis2020MNRAS} claimed that the globular clusters in DF2 are in a rotating plane, even though spatially-resolved spectroscopy of the diffuse light does not show evidence for rotation \citep{Danieli2019ApJL,Emsellem2019A&A}.
\citet{Montes2021ApJ} suggested that DF2 is a low-inclination disk galaxy, based on a photometric analysis.

While such ``tail of the distribution'' explanations can be invoked for a single unusual galaxy, the similarity of DF2 and DF4 strongly suggests a common origin. \citet{Keim2022ApJ} showed that the similarity extends even to their tidal distortions, with both galaxies showing the same ellipticity and (absolute) orientation as a function of radius. It is also unlikely that both galaxies are face-on disks.
\citet{Silk2019MNRAS} proposed the ``mini bullet-cluster'' event, where a high-speed collision of two progenitor gas-rich galaxies produced a dark-matter deficient galaxy.
During such a collision, the collision-less dark matter halos and stars would keep moving in their orbits, while the collisional gas would be separated. Shocks and gravitational collapse could then lead to the
formation of stars and globular clusters.
Following the \citet{Silk2019MNRAS} idea, \citet{vanDokkum2022Natur} suggested that DF2 and DF4 are both remnants of the same collision that occurred about 9 Gyr ago and discovered a linear trail of $\approx10$ low-surface brightness galaxies in the NGC1052 group which are thought to have formed in the aftermath of the collision. 

Given the various formation {scenarios}, the definite dark matter content of DF2 and DF4 carries important implications.
For DF2, the velocity dispersion has been independently inferred from its GCs ($7.8^{+5.2}_{-2.2}$, \citealt{vanDokkum2018Natur}; $10.5^{+4.0}_{-2.2}$, \citealt{Emsellem2019A&A}) and from the velocity dispersion of its diffuse stellar light \citep[$8.5^{+2.3}_{-3.1}$,][]{Danieli2019ApJL}. 
However, for DF4, its velocity dispersion has been measured only once with its seven GCs \citep{vanDokkum2019ApJL}, probing the enclosed mass within
$7\,\mathrm{kpc}$.

An accurate stellar velocity dispersion for DF4 would not only test whether the galaxy is indeed dark matter-deficient but also help constrain the formation of dark-matter-deficient galaxies. For example, in stripping scenarios one might expect some dark matter to remain with the galaxy, whereas in the bullet scenario the galaxy should be truly devoid of dark
matter. Distinguishing a velocity dispersion
of, say, $12$\,\kms\ from $7$\,\kms\ requires
high spectral resolution as well as high sensitivity, which is particularly difficult to achieve for low-surface-brightness galaxies.
Integral Field Unit (IFU) spectrographs can act as a ``light bucket'' to build up signal-to-noise for these faint and large galaxies \citep[e.g.][]{Danieli2019ApJL,Emsellem2019A&A,Gannon2023MNRAS}. 
The observed broadening of spectral features is $$ \sigma_{\mathrm{obs}}^2 = \sigma_{\mathrm{instrument}}^2 + \sigma_{\mathrm{galaxy}}^2. $$
Because of this quadratic behavior, a spectral resolution $$\sigma_{\mathrm{instrument}} \sim \sigma_{\mathrm{galaxy}} $$ is required.

For the expected intrinsic velocity dispersion of DF4, the Keck Cosmic Web Imager (KCWI) is the only IFU that can achieve this instrumental resolution. 
Combining its ``light bucket'' capability with the light gathering power of the Keck telescope, KCWI allows us to take the highest resolution spectrum of any low-surface-brightness galaxy. In this paper, we present KCWI spectroscopy of DF4 with an instrumental resolution of 7\kms (Section \ref{sec:instrument}). We measure the velocity dispersion of the diffuse stellar component of DF4 (Section \ref{sec:veldispersion}) and confirm that it {contains little to no meaurable dark matter.}

\section{Observation and Data Reduction}

\subsection{KCWI spectroscopy}

IFU spectroscopy of DF4 was obtained with KCWI on Keck II in October 2019 and November 2021 for a total of four nights.
The small image slicer was used with the high-resolution BH2 grating, resulting in a field-of-view (FOV) of $8\arcsec \times 20\arcsec$ {($0.8 \times 1.9\,$kpc)}
and an approximate spectral resolution of $R\sim18,000$. 
The data were taken with $1\times1$ binning and sky position angle of 0.
The central wavelength was 4250\AA\ and the wavelength coverage was $4040-4450$\AA.

NGC1052-DF4's angular size ($R_e=16.5\arcsec=1.6\,\mathrm{kpc}$) is larger than the KCWI FOV, which means that offset exposures have to be used to characterize the sky emission. 
As illustrated in Figure \ref{fig:FOV}, {we alternate between two pointings}. 
In the first, ``science'' exposures were taken with the KCWI FOV placed on the center of NGC1052-DF4. Due to dithering, the effective FOV is $12.25\arcsec \times 20.4\arcsec$ {($1.2\times 1.9\,$kpc)}, 
covering the inner 95\% $R_e$ of the diffuse stellar component, as well as GC-1968 and GC-2239, globular clusters associated with the galaxy. 
In the second pointing, offset ``sky'' exposures were taken with the FOV placed on a field $41.6\arcsec$ away, centered on the globular cluster GC-2537. 
The globular clusters take up only a small fraction of the KCWI area, and are masked in the analysis presented in this paper. 
{In the data reduction process (described in Section \ref{sec:reduction}), sky subtraction was done separately for each epoch.}

In this high resolution mode the read noise of the KCWI detector exceeds the sky noise in short exposures. The exposure time is therefore a compromise between getting as close to sky-limited data as possible, while still having enough {number of science and sky exposures} to perform the data reduction.
We chose exposure times of 3600\,s at each position, the approximate {transition} point where the typical sky noise begins to exceed the read noise.
The total exposure time is
50,400\,s on the galaxy and 32,400\,s on the offset field. 
The total science + sky time that is used in the analysis is
therefore 34 hours. 
Conditions were somewhat variable, with thin cirrus present during 2019 and clear skies in 2021.

\subsection{Data Reduction} \label{sec:reduction}
The  KCWI Data Reduction Pipeline (KCWI DRP) was used to perform basic reduction and calibration of the data
\footnote{\url{https://github.com/Keck-DataReductionPipelines/KCWI_DRP}}.
A background gradient was manually removed from the master dark created by the KCWI DRP.
A 9th-order polynomial was fitted to each row and the average fit was subtracted from the master dark.
In addition to the master dark, the event table was also edited to skip the default sky subtraction. 
The rest of the data reduction procedures were set to their defaults.
The KCWI DRP treats each of the 34 science and sky frames independently. 
{``Bars'' exposures are used to identify geometric transformations required to map each pixel in the 2D image into
slice and position; the arc lamp exposures provide the wavelength solution.}
These transformations were used to convert the raw science image into a data cube with three dimensions: slice number, position along the slice, and wavelength. 
These data cubes, dubbed \texttt{icubed} files by the KCWI DRP but without sky subtraction, were used in the subsequent steps.

Sky subtraction was carried out separately with a custom method that is optimized for low-surface-brightness galaxies. 
The sky frames could not be directly subtracted from adjacent science frames because the sky spectrum varied significantly over the hour-long interval between successive exposures. 
Instead, variation in the sky spectrum was modelled with a principal component analysis (PCA). The method was introduced and explained in detail in \citet{vanDokkum2019ApJ} and was also used for KCWI spectroscopy of NGC1052-DF2 \citep{Danieli2019ApJL}. 

To summarize the procedure, principle components extracted from the sky spectra were used as templates to fit the science frames. The sky data set included both the offset field and empty field exposures at the beginning and end of each night. All sky frames were masked to exclude GCs and bad pixels, then averaged over the two spatial dimensions.
Due to the low signal-to-noise in each pixel, it was essential to discard sky pixels that deviate $\geq 5 \sigma$ from the median before taking the average.
The resulting 1D sky spectra were grouped by observing epoch (2019 and 2021) and separately decomposed into eight components with \texttt{scikit-learn} PCA. 
For each epoch, a fitting template was compiled from the eigenspectra, the epoch-average 1D sky spectra, and a model of the galaxy spectrum.
These templates were fitted to each science exposure in the same epoch with their amplitudes as fitting parameters. The best-fit 1D sky model was subtracted from each spatial pixel. We refer to \citet{vanDokkum2019ApJ} for details.

\begin{figure*}[htbp!]
\plotone{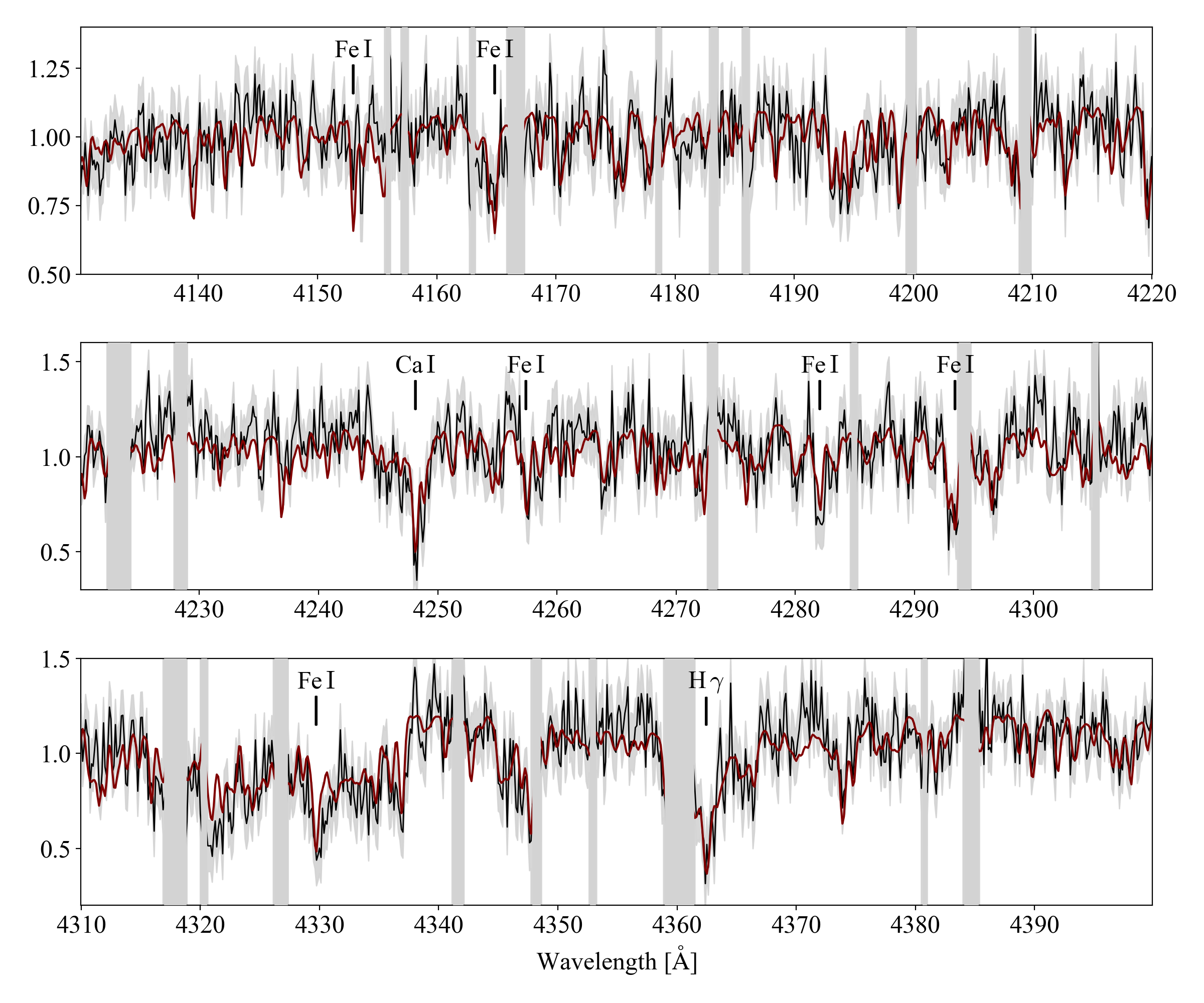}
\caption{Integrated 14hr KCWI spectrum of NGC1052-DF4 (black line) with instrumental resolution of 0.107\AA. The high spectral resolution reveals a large number of absorption lines with high accuracy. The $1\sigma$ uncertainties and masked regions are shaded in grey. The best-fit model from synthetic stellar populations is over-plotted in red.}
\label{fig:spec}
\end{figure*}

After sky subtraction a 1D spectrum was extracted from each science datacube. Again, the pixels were averaged over and any pixels that deviated $\geq 5 \sigma$ from the median are not included in the 1D average spectra.
The final spectrum is a weighted average of the 14 science spectra, where the weights were determined from the integrated GC flux of each datacube. 
The final combined spectrum is shown in Fig \ref{fig:spec}.

\section{Measuring Stellar Kinematics}
\label{sec:instrument}

\subsection{Instrumental Resolution}
Since the instrumental resolution contributes to the stellar velocity dispersion measured from the spectrum, it is important to determine it as a function of wavelength.
The instrumental resolution is typically determined from the width of emission lines from the ThAr arc lamp. The median width of the arc lines is $0.115 \pm 0.003$ \AA \,in 2019, and $0.118 \pm 0.002$ \AA\, in 2021. The arc frames show stable instrumental resolution between our two observing epochs. 

{The effective spectral resolution of the science data could differ from the arc frame resolution due to several reasons: arc frames are only taken at the beginning of night, but the science and sky frames span the whole night; the light path from the arc lamp to the detector is different from the path of the galaxy light; multiple science frames are combined in the final analysis, which can also impact the effective spectral resolution.} A {more empirical} approach is to measure the instrumental resolution directly from the reduced {sky} data, {using the same method that we measure velocity dispersion from the reduced science data. }

Similar to \citet{vanDokkum2019ApJ}, we make use of the fact that the first eigenspectra (PC1) of our 2019 data shows scattered and reflected sunlight. The instrumental resolution is determined by fitting a very high resolution solar template to PC1, with the instrumental broadening a free parameter. 
The high-resolution solar template is obtained from the BAse de données Solaire Sol (BASS2000\footnote{\url{http://bass2000.obspm.fr/solar_spect.php}}). 
The template and PC1 are split into chunks of 50\AA\, and each chunk is fitted separately.
Free parameters in the fit are the systemic velocity, the velocity dispersion, an additive constant, and a multiplicative constant.
There is good correspondence between PC1 and the solar template, and the average instrumental resolution over the entire wavelength range is 0.107 \AA.
The measured resolution from each chunk stays within 0.003\AA\, of the average so we will assume a constant instrumental resolution as a function of wavelength.
Since this method encapsulates the effective spectral resolution, we use $\sigma_{\mathrm{instrumental}} = 0.107$\,\AA\ to construct the template for the science spectrum. Using the resolution from the arc lines instead leads to 0.6\kms\ smaller final dispersions. 

\begin{figure}[htbp!]
    \centering
    \plotone{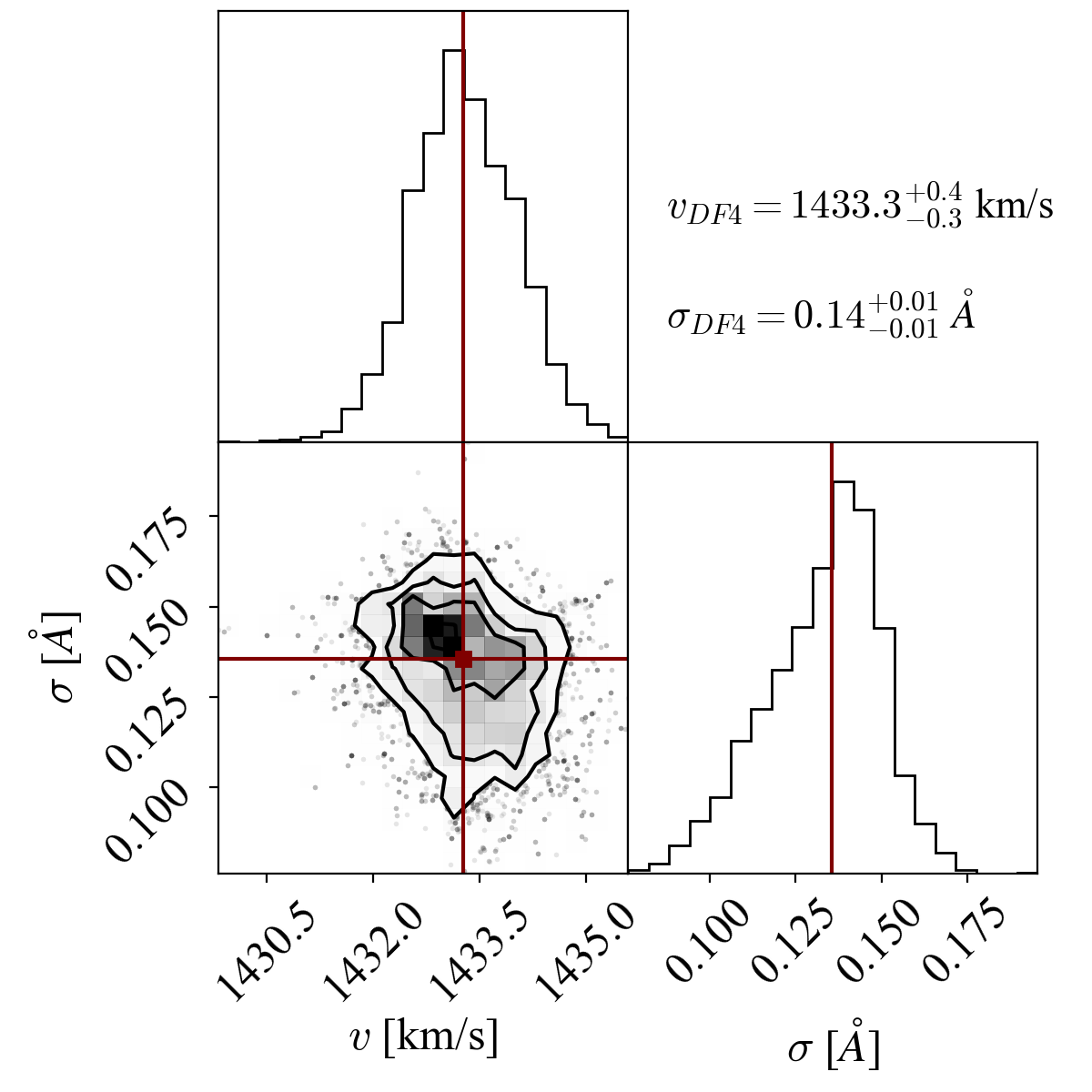}
    \caption{The posterior distribution of the systemic velocity ($v$) and line width ($\sigma$) from MCMC fitting of the KCWI spectrum.}
    \label{fig:mcmc}
\end{figure}

\subsection{Line width}
The width of the absorption lines of NGC1052-DF4 is measured by fitting template spectra to the combined 1D spectrum. The templates come from a set of
synthetic stellar population synthesis models with an intrinsic wavelength resolution of 0.04\AA\
\citep{Conroy2018ApJ}.
No spectroscopic metallicity has so far been measured for the diffuse light of DF4.
A photometric measurement, and spectroscopic measurements for the globular clusters and diffuse light in DF2, indicate metallicities
of $-1.5 \lesssim [Z/H]\lesssim -1.0$ and
ages of $7-9$\,Gyr \citep{vanDokkum2018Natur,Fensch2019A&A,Buzzo2022MNRAS}. 
We used two templates with metallicity $[Z/H]=-1.5$ and different ages: 9 Gyrs and 5 Gyrs. 
Given the age -- metallicity degeneracy, the
younger model is equivalent to an older model with a higher metallicity.
The templates were convolved to the KCWI resolution and supplied to the MCMC fitting code.

We fit the final combined spectrum in the wavelength region 4130\AA $\leq \lambda \leq$ 4400\AA, excluding the region between 4050\AA $\leq \lambda \leq$ 4130\AA. A strong feature in the sky spectrum in this wavelength range caused a mismatch between the observed spectrum and the expected H$\delta$ line, so we begin the fit at 4130\AA.

The velocity and velocity dispersion were determined using an MCMC methodology first described in \citet{vanDokkum2016ApJL}.
The algorithm finds the best linear combination of the two templates while using a seventh-order multiplicative polynomial and a second-order additive polynomial to account for the continuum. 
Flat priors are used for each fitting parameter, including the redshift (0.00478) and line width ($0< \sigma < 0.5$ \AA).
We iteratively identify outliers and mask the regions that deviate significantly from the model.
This was necessary because residual instrumental effects and sky lines (e.g. around 4360\AA) affect a small fraction of the final spectrum. 
The best-fitting model is shown in Figure \ref{fig:spec} as the red line and the masked intervals are shaded in grey. 
The reduced chi-square value of the best-fit model is 0.98 and the MCMC parameters converge. The posterior distribution of the systemic velocity $v_{\rm sys}$ and the velocity dispersion $\sigma$ is shown in Figure \ref{fig:mcmc}.

From the high-resolution KCWI spectra of the diffuse stellar light in DF4, 
we measure a line width of $\sigma_{\mathrm{galaxy}}=9.67_{-0.9}^{+1.4}$\kms. This value includes both stellar velocity dispersion and stellar rotation.

\section{Results}
\label{sec:veldispersion}
\begin{figure}[htbp!]
    \centering
    \plotone{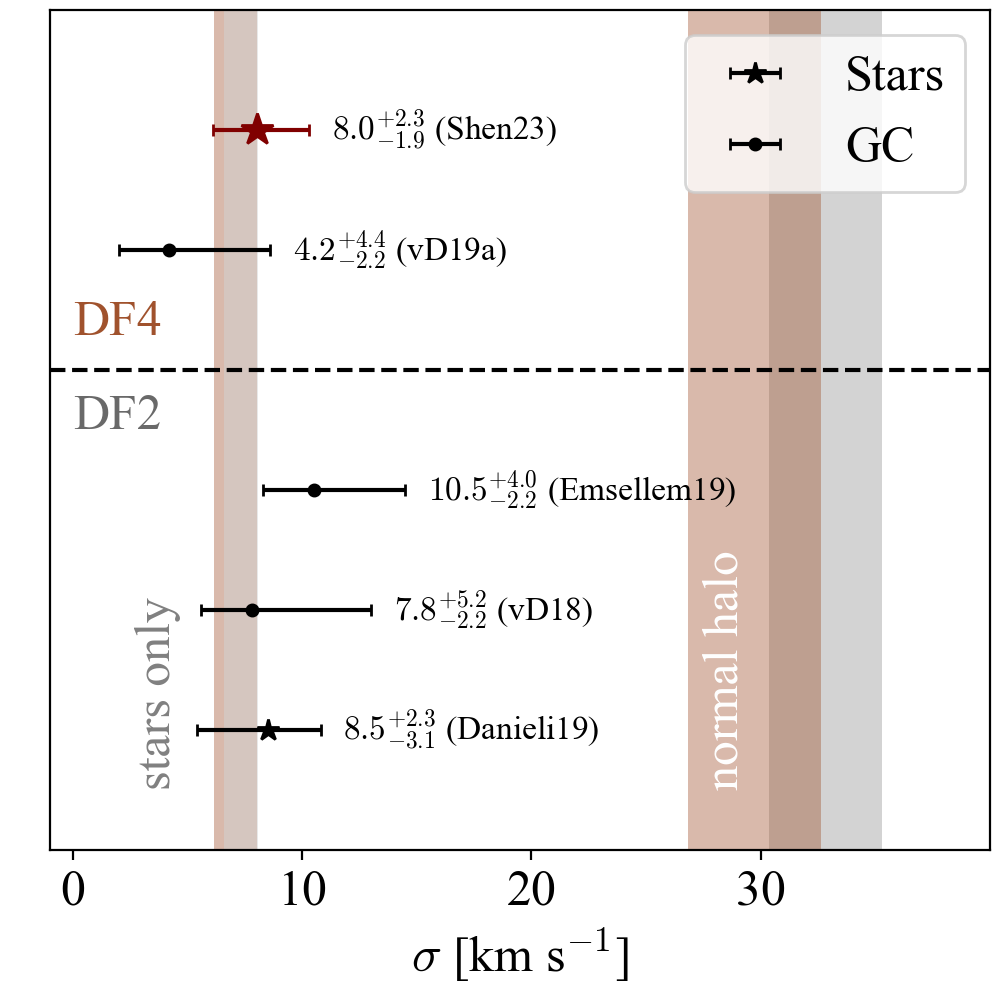}
    \caption{The best-fit velocity dispersion from this study  (red star) corrected for stellar rotation and compared to literature values. For DF4, the stars alone contribute $8 \sim 10$\,\kms (light brown band) while the expectation from the stellar-mass-halo-mass relation \citep{Zaritsky2023MNRAS}, assuming a standard NFW halo, is $\sigma_{\mathrm{SMHM}} = 30\pm5$\,\kms. The stellar contribution of DF2 is shaded in grey bands.}
    \label{fig:comparison}
\end{figure}

\subsection{Stellar rotation}
The integrated spectrum of diffuse stellar light in a galaxy is broadened both by velocity dispersion between stars and internal motion of individual stars (largely rotation and macroturbulence): $$\sigma_{\mathrm{galaxy}}^2=\sigma_{\mathrm{stars}}^2+\sigma_{\mathrm{broadening}}^2$$ For galaxies with high dark matter content, the velocity dispersion dominates. In dark-matter-deficient galaxies like DF4, the velocity dispersion is so low ($< 10 $\kms) that the line broadening of individual stars may have a noticeable effect on the measured line width \citep{Carney2008AJ,Massarotti2008AJ}. Since our synthetic stellar population synthesis template does not include intrinsic broadening, we need to estimate it separately to measure the stellar velocity dispersion. 

The stellar population in DF4 is old ($\sim9\,\mathrm{Gyr}$) and metal-poor \citep{Buzzo2022MNRAS}, mostly composed of main sequence (MS) stars and red giant branch (RGB) stars.
The internal motions of metal-poor stars are not well-determined in the literature. Sun-like MS stars have rotation velocities of around 3\,\kms \citep{Sheminova2019KPCB}, but metal-poor MS stars have been predicted to have rotation velocities as high as 6\kms \citep{Amard2020ApJ}. When a MS star evolves to the RGB, its rotation rate decreases (known as spin-down), but the rate of spin down depends on metallicity. Furthermore, metal-poor red giants can have high macroturbulence \citep[6-8\kms, ][]{Gray1982ApJ}, an additional source of spectral line broadening.
Observationally, \citet{Carney2008AJ} and \citet{Massarotti2008AJ} measured  internal broadening of 2--11\,\kms\ in RGB stars. From the observed metallicity vs.\ broadening relation in \citet{Carney2008AJ}, we infer an internal broadening of $\sigma_{\mathrm{broadening}}=5.4 \pm2.0$\kms\ in red giants with $[Fe/H]=-1$. The $\pm 1\sigma$
range 3.4 -- 7.4\,\kms\ encompasses the plausible macroturbulence in red giants, as well as the broadening due to rotation of Sun-like stars at the low end up to low metallicity MS stars at the high end.

\subsection{Velocity dispersion}
After correcting for this internal broadening of $5.4 \pm2$\kms, we obtain a velocity dispersion of   $\sigma_{\mathrm{stars}}=8.0_{-1.9}^{+2.3}$\kms.
In Fig.\ \ref{fig:comparison} we summarize the velocity dispersion measurements for DF2 and DF4, including the new value of the stellar dispersion. 
The exceptional spectral resolution of KCWI (7 \kms using the small slicer and the BH2 grating) produces the smallest errorbar of the five values.

The only previous velocity dispersion measurement of DF4 came from 7 globular clusters \citep{vanDokkum2019ApJL}. Analyzing such a small number of dynamical tracers with regular MCMC could lead to a biased dispersion \citep{Laporte2019MNRAS}, and recent papers have proposed modifying MCMC priors for an unbiased estimate \citep{Doppel2021MNRAS,Toloba2023ApJ}. For the velocity dispersion of DF4, \citet{vanDokkum2019ApJL} used approximate Bayesian computation, which does not rely on explicit likelihood or prior functions and does not suffer from the small sample bias. The globular cluster dispersion of $\sigma_{\mathrm{GC}}=4.2_{-2.2}^{+4.4}$\kms covers a median galactocentric radius of 4.08 kpc ($2.6\,R_{e}$). Our new KCWI velocity dispersion probes the inner diffuse light out to an average radius of $\sim0.5\, R_{e}$. These two measurements are not sufficient to constrain the radial velocity dispersion profile, so we 
combine them with a simple weighted average to obtain a fiducial velocity dispersion of $\sigma_{\rm f} = 6.3_{-1.6}^{+2.5}$\,\kms\ for DF4.

\section{Discussion}

We have used KCWI on the Keck\,II telescope to measure the stellar velocity dispersion of DF4, the ``second galaxy lacking dark matter'' in the NGC1052 group. Combining the stellar dispersion with the previously-measured dispersion from the velocities of globular clusters, we find a fiducial value of $\sigma_{\rm f} = 6.3_{-1.6}^{+2.5}$\,\kms\ for DF4.

We calculate the dynamical mass of DF4 within the half-light radius, using the \citet{Wolf2010MNRAS} relation:
\begin{equation}
M_{1/2} = 3\, G^{-1} \langle \sigma_{\mathrm{LOS}}^2 \rangle\, R_{1/2}
\end{equation}
The projected circularized {half-light radius of $R_e=1.6 \pm 0.1$ kpc \citep{Cohen2018ApJ, Danieli2020ApJL}} translates to a 3D half-light radius $R_{\mathrm{1/2}}\approx 4/3 R_e \approx 2.1 \pm 0.1$ kpc. Using the fiducial disersion we find $M(r<R_{1/2})= 8_{-4}^{+6}  \times 10^7 M_{\odot}$ within this radius. 

This mass is fully consistent with
the stellar mass within {the half-light radius of $M_*(r<R_{1/2}) = (7.5 \pm 2) \times 10^7 M_{\odot}$ \citep[assuming $M/L= (2.0\pm 0.5) M_{\odot}/L_{\odot}$, ][]{vanDokkum2019ApJL}}, and we infer that our results are consistent with dark matter-free models. We can also rule out a standard NFW halo that follows the
stellar mass -- halo mass relation \citep{Behroozi2013ApJ}: the expectation is $M_{\mathrm{NFW}}(r<r_{1/2})= 1.2\times 10^9 M_{\odot}$, orders of magnitude higher than the dynamical mass of DF4. {More recent works on the satellite stellar mass -- halo mass relation \citep{Nadler2020ApJ, Danieli2022arXiv} predict even higher halo masses at this stellar mass, exacerbating the discrepancy.}

{DF4 is an outlier not only in the stellar mass -- halo mass relation, but also in the GC system -- halo mass relation. Massive galaxies follow the simple linear relationship $M_h = 5\times 10^9 M_{\odot}\, N_{GC}$, which can also be expressed in the GC mass instead of their number  \citep{Spitler2009MNRAS, Forbes2018MNRAS, Burkert2020AJ}. In the dwarf galaxy regime ($10^{8}<M_h< 10^{10} M_{\odot}$), this relation only applies in an average sense because the predicted number of GCs is small and the nature of GC formation is stochastic.}
{Both DF2 and DF4 are consistent with having no dark matter based on the kinematics of their stars and GC system, but they each host $19^{+9}_{-5}$ GCs \citep{Shen2021ApJ} which are overmassive by a factor of 4. Using the canonical relation, and multiplying by a factor of four to account for the high mass of individual clusters, we find an expected halo mass of $M_h = 4^{+2}_{-1} \times 10^{11}  M_{\odot}$. For an NFW halo with this mass, the dark matter mass enclosed within $R_{1/2}$ is $5^{+3}_{-2} \times 10^9 M_{\odot}$, between 10 and 30 times higher than our $3-\sigma$ upper limit on the dynamical mass of DF4.}

{Three factors contribute to the outlier status of DF2 and DF4: they have less dark matter than expected for their stellar mass, more GCs than expected for their stellar mass, and a higher mass of individual clusters than other galaxies. 
The outlier status in the GC system -- halo mass relation is a different way to
express the potentially exotic origin of the galaxies. The GCs in DF2 and DF4 do not follow the near-universal GC luminosity function \citep{Shen2021ApJ}, but they are remarkably consistent in color \citep{vanDokkum2022ApJL}. If these GCs did not follow the canonical GC formation or hierarchical accretion, then the GC mass -- halo mass relation may not apply here.
} 

Interestingly, the fiducial dispersion is in some
tension with cored dark matter profiles, which might result from extreme tidal stripping. \citet{Carleton2019MNRAS} predict a velocity
dispersion of 
$\sim10$\kms at $R_{e}=1.6$\,kpc,  1.5$\sigma$ higher than the fiducial dispersion. It is {clearly inconsistent} with the MOND prediction of $\geq 12.5$\,\kms\ \citep{Muller2019A&A_mond}. This is a strict lower limit in MOND, as corresponds to the maximum external gravitational field of NGC1052.

It is difficult to improve the accuracy further.
Systematic uncertainties in the contribution from intrinsic stellar broadening, as well as the stellar initial mass function, are at the same level as the random uncertainty. A more fruitful approach may be to expanding the sample. In the ``bullet dwarf'' model there are several other candidates for dark matter-free objects in the NGC1052 group. The kinematics of these galaxies will provide further constraints on formation scenarios and the nature of dark matter-deficient galaxies.


\begin{acknowledgments}
We thank Imad Pasha and Michael Keim for their help during the Keck observations. Support from STScI grant HST-GO-15695 is gratefully acknowledged.
\end{acknowledgments}

\vspace{5mm}
\facilities{HST(ACS), Keck(KCWI)}


\software{astropy \citep{AstropyCollaboration2013A&A, AstropyCollaboration2018AJ, AstropyCollaboration2022ApJ}
          }

\appendix

\section{A Small Revision to the TRGB magnitude of NGC1052-DF2}\label{appdx}

Along with the velocity dispersions of both galaxies,
the line-of-sight distance between DF2 and DF4 is an important constraint on models for their formation.
Here we report a small correction to the published distance.
\citet{Shen2021ApJL} used 40 orbits of HST data to find the Tip-of-the-Red-Giant-Branch (TRGB) distance to NGC1052-DF2. The 2021 analysis found a TRGB magnitude of $m_{TRGB, F814W} = 27.52$ mag, corresponding to an absolute distance of 22.1 Mpc and a distance between DF2 and DF4 of $2.1\pm 0.5$\,Mpc.

In the original 2021 data analysis, two out of the 20 orbits in each filter were misaligned with the rest. This error occurred during the \texttt{tweakreg} routine of the \texttt{AstroDrizzle} package, which identifies stars in each raw image and finds the best shift to align them before the drizzle combination process. Due to the low surface brightness of our target galaxy, \texttt{tweakreg} failed to align two out of 20 orbits in each filter. Due to the offset (around 10 pixels), these two orbits were effectively rejected during the final \texttt{drizzle} process and did not leave artifacts in the combined image. We found this error because the weight map showed pixels preferentially rejected at the location of bright objects like globular clusters.
This misalignment also affected the Dolphot photometry because it relied on the WCS header from each \texttt{flc} file. Since \texttt{tweakreg} failed for two orbits and the shift was larger than the tolerance of Dolphot, Dolphot missed the flux of all the stars in the misaligned images. Around 10\% of our data were misaligned, and it should lead to around 0.1 mag underestimation of the TRGB location. 
\begin{figure}[htbp!]
    \centering
    \plottwo{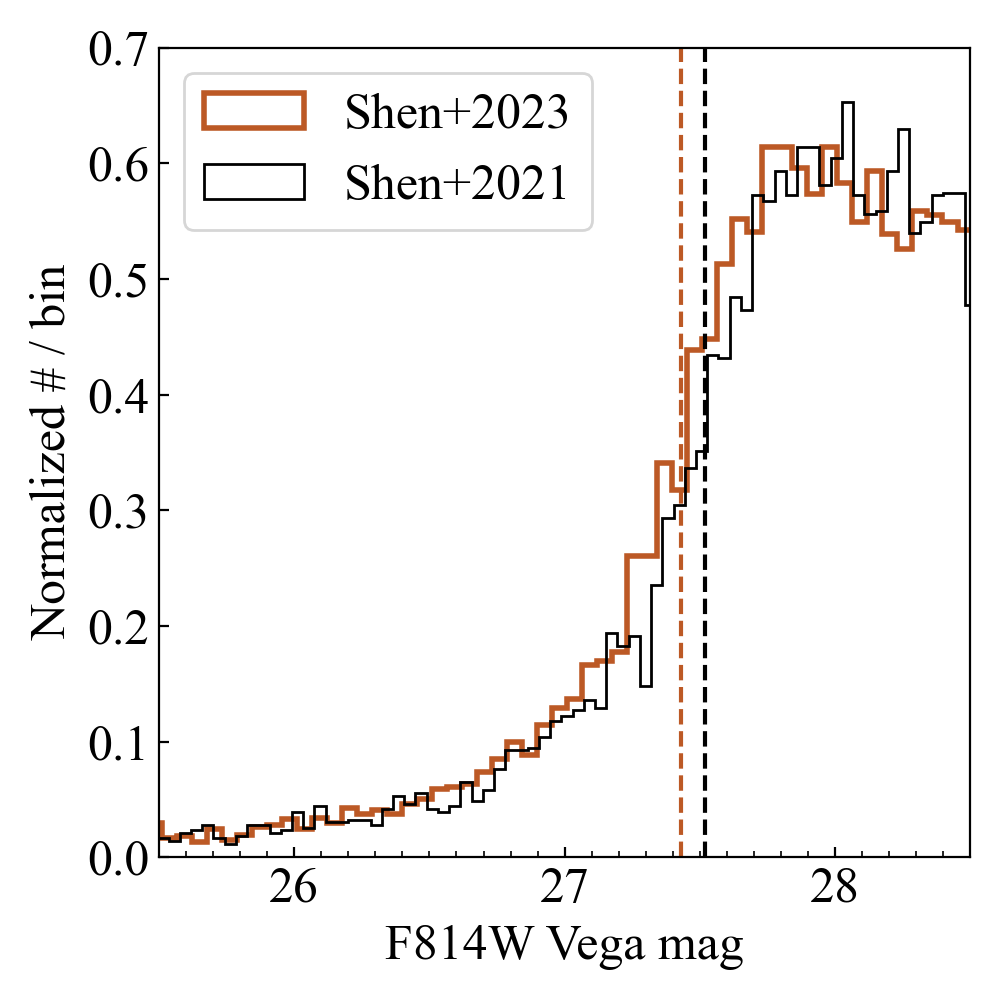}{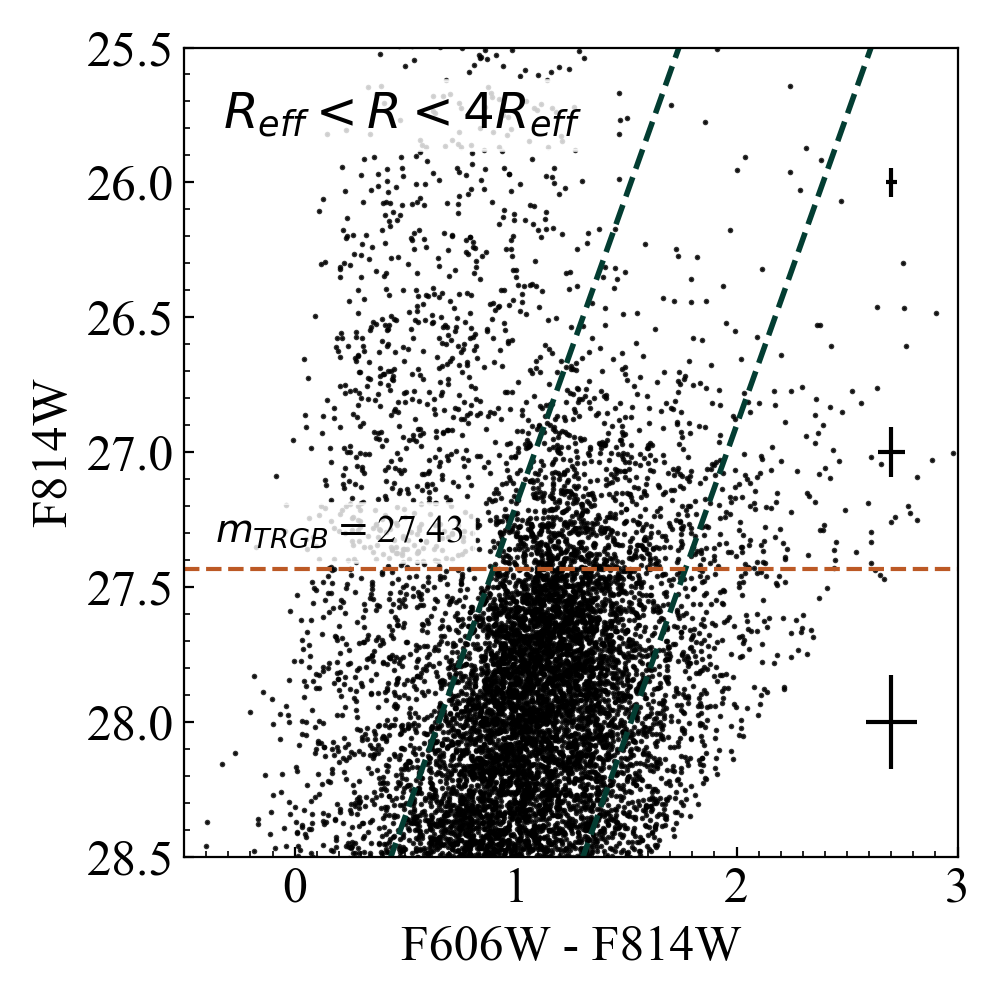}
    \caption{Left panel: the observed F814W luminosity function of the Red Giant Branch in NGC1052-DF2. The solid orange line shows the luminosity function from the revised analysis, while the solid black line shows the luminosity function as published in \citet{Shen2021ApJL}. The dashed lines indicate the location of steepest increase in both luminosity functions: 27.43 mag for the 2023 reanalysis, and 27.52 mag for the \citet{Shen2021ApJL} analysis. Right panel: the revised color-magnitude diagram of NGC1052-DF2. The error bars come from artificial star tests, and the dark green dashed lines indicate the location of the Red Giant Branch.}
    \label{fig:lf_21_22}
\end{figure}

We re-analyzed the HST data by dropping the two misaligned orbits from the Dolphot input, leaving everything else the same. From a simple edge detection of the RGB luminosity function (shown in the left panel of Figure \ref{fig:lf_21_22}), we revise the observed TRGB magnitude from 27.52 mag to 27.43 mag. The 0.1 mag shift we measure can be explained by the fraction of orbits missed.

The re-run of 180,000 artificial stars show that the remaining 18 orbits is still deep enough to reach below the TRGB, so we proceed with re-running the forward modeling analysis. Our forward modeling code starts with an intrinsic luminosity function and takes into account the photometric bias and scatter measured from the artificial stars. Compared to the 2021 value, the new TRGB magnitude is brighter, leading to lower photometric bias levels. This partially compensates for the change in the observed TRGB magnitude. From the forward modeling, we find the intrinsic TRGB magnitude of DF2 to be $m_{\mathrm{F814W, TRGB}} = 27.61$ mag, which is 0.06 mag brighter than the 2021 measurement.

The revised distance we measure from the forward modeling is 21.7\,Mpc (from 22.1\,Mpc). The
revised distance between DF2 and DF4 is $1.7\pm 0.5$\,Mpc.

\begin{figure*}
    \centering
    \includegraphics[width=0.9\textwidth]{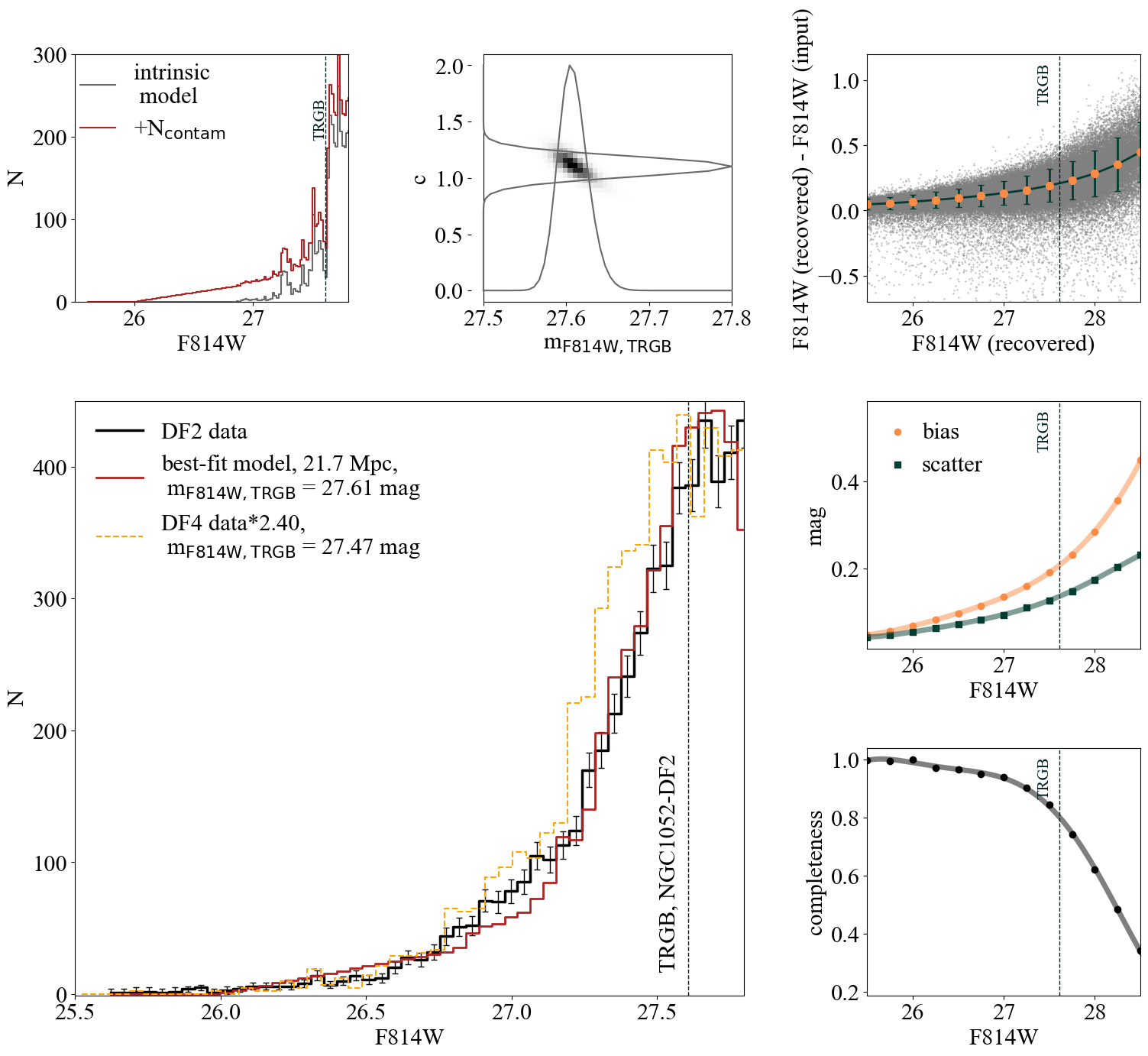}
    \caption{Re-analysis of the TRGB forward modeling.}
    \label{fig:TRGB_22}
\end{figure*}


\bibliography{paper}{}

\begin{thebibliography}{}
\expandafter\ifx\csname natexlab\endcsname\relax\def\natexlab#1{#1}\fi
\providecommand{\url}[1]{\href{#1}{#1}}
\providecommand{\dodoi}[1]{doi:~\href{http://doi.org/#1}{\nolinkurl{#1}}}
\providecommand{\doeprint}[1]{\href{http://ascl.net/#1}{\nolinkurl{http://ascl.net/#1}}}
\providecommand{\doarXiv}[1]{\href{https://arxiv.org/abs/#1}{\nolinkurl{https://arxiv.org/abs/#1}}}

\bibitem[{{Amard} \& {Matt}(2020)}]{Amard2020ApJ}
{Amard}, L., \& {Matt}, S.~P. 2020, \apj, 889, 108,
  \dodoi{10.3847/1538-4357/ab6173}

\bibitem[{{Astropy Collaboration} {et~al.}(2013){Astropy Collaboration},
  {Robitaille}, {Tollerud}, {Greenfield}, {Droettboom}, {Bray}, {Aldcroft},
  {Davis}, {Ginsburg}, {Price-Whelan}, {Kerzendorf}, {Conley}, {Crighton},
  {Barbary}, {Muna}, {Ferguson}, {Grollier}, {Parikh}, {Nair}, {Unther},
  {Deil}, {Woillez}, {Conseil}, {Kramer}, {Turner}, {Singer}, {Fox}, {Weaver},
  {Zabalza}, {Edwards}, {Azalee Bostroem}, {Burke}, {Casey}, {Crawford},
  {Dencheva}, {Ely}, {Jenness}, {Labrie}, {Lim}, {Pierfederici}, {Pontzen},
  {Ptak}, {Refsdal}, {Servillat}, \& {Streicher}}]{AstropyCollaboration2013A&A}
{Astropy Collaboration}, {Robitaille}, T.~P., {Tollerud}, E.~J., {et~al.} 2013,
  \aap, 558, A33, \dodoi{10.1051/0004-6361/201322068}

\bibitem[{{Astropy Collaboration} {et~al.}(2018){Astropy Collaboration},
  {Price-Whelan}, {Sip{\H{o}}cz}, {G{\"u}nther}, {Lim}, {Crawford}, {Conseil},
  {Shupe}, {Craig}, {Dencheva}, {Ginsburg}, {VanderPlas}, {Bradley},
  {P{\'e}rez-Su{\'a}rez}, {de Val-Borro}, {Aldcroft}, {Cruz}, {Robitaille},
  {Tollerud}, {Ardelean}, {Babej}, {Bach}, {Bachetti}, {Bakanov}, {Bamford},
  {Barentsen}, {Barmby}, {Baumbach}, {Berry}, {Biscani}, {Boquien}, {Bostroem},
  {Bouma}, {Brammer}, {Bray}, {Breytenbach}, {Buddelmeijer}, {Burke},
  {Calderone}, {Cano Rodr{\'\i}guez}, {Cara}, {Cardoso}, {Cheedella}, {Copin},
  {Corrales}, {Crichton}, {D'Avella}, {Deil}, {Depagne}, {Dietrich}, {Donath},
  {Droettboom}, {Earl}, {Erben}, {Fabbro}, {Ferreira}, {Finethy}, {Fox},
  {Garrison}, {Gibbons}, {Goldstein}, {Gommers}, {Greco}, {Greenfield},
  {Groener}, {Grollier}, {Hagen}, {Hirst}, {Homeier}, {Horton}, {Hosseinzadeh},
  {Hu}, {Hunkeler}, {Ivezi{\'c}}, {Jain}, {Jenness}, {Kanarek}, {Kendrew},
  {Kern}, {Kerzendorf}, {Khvalko}, {King}, {Kirkby}, {Kulkarni}, {Kumar},
  {Lee}, {Lenz}, {Littlefair}, {Ma}, {Macleod}, {Mastropietro}, {McCully},
  {Montagnac}, {Morris}, {Mueller}, {Mumford}, {Muna}, {Murphy}, {Nelson},
  {Nguyen}, {Ninan}, {N{\"o}the}, {Ogaz}, {Oh}, {Parejko}, {Parley}, {Pascual},
  {Patil}, {Patil}, {Plunkett}, {Prochaska}, {Rastogi}, {Reddy Janga},
  {Sabater}, {Sakurikar}, {Seifert}, {Sherbert}, {Sherwood-Taylor}, {Shih},
  {Sick}, {Silbiger}, {Singanamalla}, {Singer}, {Sladen}, {Sooley},
  {Sornarajah}, {Streicher}, {Teuben}, {Thomas}, {Tremblay}, {Turner},
  {Terr{\'o}n}, {van Kerkwijk}, {de la Vega}, {Watkins}, {Weaver}, {Whitmore},
  {Woillez}, {Zabalza}, \& {Astropy Contributors}}]{AstropyCollaboration2018AJ}
{Astropy Collaboration}, {Price-Whelan}, A.~M., {Sip{\H{o}}cz}, B.~M., {et~al.}
  2018, \aj, 156, 123, \dodoi{10.3847/1538-3881/aabc4f}

\bibitem[{{Astropy Collaboration} {et~al.}(2022){Astropy Collaboration},
  {Price-Whelan}, {Lim}, {Earl}, {Starkman}, {Bradley}, {Shupe}, {Patil},
  {Corrales}, {Brasseur}, {N{\"o}the}, {Donath}, {Tollerud}, {Morris},
  {Ginsburg}, {Vaher}, {Weaver}, {Tocknell}, {Jamieson}, {van Kerkwijk},
  {Robitaille}, {Merry}, {Bachetti}, {G{\"u}nther}, {Aldcroft},
  {Alvarado-Montes}, {Archibald}, {B{\'o}di}, {Bapat}, {Barentsen},
  {Baz{\'a}n}, {Biswas}, {Boquien}, {Burke}, {Cara}, {Cara}, {Conroy},
  {Conseil}, {Craig}, {Cross}, {Cruz}, {D'Eugenio}, {Dencheva}, {Devillepoix},
  {Dietrich}, {Eigenbrot}, {Erben}, {Ferreira}, {Foreman-Mackey}, {Fox},
  {Freij}, {Garg}, {Geda}, {Glattly}, {Gondhalekar}, {Gordon}, {Grant},
  {Greenfield}, {Groener}, {Guest}, {Gurovich}, {Handberg}, {Hart},
  {Hatfield-Dodds}, {Homeier}, {Hosseinzadeh}, {Jenness}, {Jones}, {Joseph},
  {Kalmbach}, {Karamehmetoglu}, {Ka{\l}uszy{\'n}ski}, {Kelley}, {Kern},
  {Kerzendorf}, {Koch}, {Kulumani}, {Lee}, {Ly}, {Ma}, {MacBride}, {Maljaars},
  {Muna}, {Murphy}, {Norman}, {O'Steen}, {Oman}, {Pacifici}, {Pascual},
  {Pascual-Granado}, {Patil}, {Perren}, {Pickering}, {Rastogi}, {Roulston},
  {Ryan}, {Rykoff}, {Sabater}, {Sakurikar}, {Salgado}, {Sanghi}, {Saunders},
  {Savchenko}, {Schwardt}, {Seifert-Eckert}, {Shih}, {Jain}, {Shukla}, {Sick},
  {Simpson}, {Singanamalla}, {Singer}, {Singhal}, {Sinha}, {Sip{\H{o}}cz},
  {Spitler}, {Stansby}, {Streicher}, {{\v{S}}umak}, {Swinbank}, {Taranu},
  {Tewary}, {Tremblay}, {de Val-Borro}, {Van Kooten}, {Vasovi{\'c}}, {Verma},
  {de Miranda Cardoso}, {Williams}, {Wilson}, {Winkel}, {Wood-Vasey}, {Xue},
  {Yoachim}, {Zhang}, {Zonca}, \& {Astropy Project
  Contributors}}]{AstropyCollaboration2022ApJ}
{Astropy Collaboration}, {Price-Whelan}, A.~M., {Lim}, P.~L., {et~al.} 2022,
  \apj, 935, 167, \dodoi{10.3847/1538-4357/ac7c74}

\bibitem[{{Behroozi} {et~al.}(2013){Behroozi}, {Wechsler}, \&
  {Conroy}}]{Behroozi2013ApJ}
{Behroozi}, P.~S., {Wechsler}, R.~H., \& {Conroy}, C. 2013, \apj, 770, 57,
  \dodoi{10.1088/0004-637X/770/1/57}

\bibitem[{{Burkert} \& {Forbes}(2020)}]{Burkert2020AJ}
{Burkert}, A., \& {Forbes}, D.~A. 2020, \aj, 159, 56,
  \dodoi{10.3847/1538-3881/ab5b0e}

\bibitem[{{Buzzo} {et~al.}(2022){Buzzo}, {Forbes}, {Brodie}, {Romanowsky},
  {Cluver}, {Jarrett}, {Laine}, {Couch}, {Gannon}, {Ferr{\'e}-Mateu}, \&
  {Okabe}}]{Buzzo2022MNRAS}
{Buzzo}, M.~L., {Forbes}, D.~A., {Brodie}, J.~P., {et~al.} 2022, \mnras, 517,
  2231, \dodoi{10.1093/mnras/stac2442}

\bibitem[{{Carleton} {et~al.}(2019){Carleton}, {Errani}, {Cooper},
  {Kaplinghat}, {Pe{\~n}arrubia}, \& {Guo}}]{Carleton2019MNRAS}
{Carleton}, T., {Errani}, R., {Cooper}, M., {et~al.} 2019, \mnras, 485, 382,
  \dodoi{10.1093/mnras/stz383}

\bibitem[{{Carney} {et~al.}(2008){Carney}, {Gray}, {Yong}, {Latham}, {Manset},
  {Zelman}, \& {Laird}}]{Carney2008AJ}
{Carney}, B.~W., {Gray}, D.~F., {Yong}, D., {et~al.} 2008, \aj, 135, 892,
  \dodoi{10.1088/0004-6256/135/3/892}

\bibitem[{{Cohen} {et~al.}(2018){Cohen}, {van Dokkum}, {Danieli}, {Romanowsky},
  {Abraham}, {Merritt}, {Zhang}, {Mowla}, {Kruijssen}, {Conroy}, \&
  {Wasserman}}]{Cohen2018ApJ}
{Cohen}, Y., {van Dokkum}, P., {Danieli}, S., {et~al.} 2018, \apj, 868, 96,
  \dodoi{10.3847/1538-4357/aae7c8}

\bibitem[{{Conroy} {et~al.}(2018){Conroy}, {Villaume}, {van Dokkum}, \&
  {Lind}}]{Conroy2018ApJ}
{Conroy}, C., {Villaume}, A., {van Dokkum}, P.~G., \& {Lind}, K. 2018, \apj,
  854, 139, \dodoi{10.3847/1538-4357/aaab49}

\bibitem[{{Danieli} {et~al.}(2022){Danieli}, {Greene}, {Carlsten}, {Jiang},
  {Beaton}, \& {Goulding}}]{Danieli2022arXiv}
{Danieli}, S., {Greene}, J.~E., {Carlsten}, S., {et~al.} 2022, arXiv e-prints,
  arXiv:2210.14233, \dodoi{10.48550/arXiv.2210.14233}

\bibitem[{{Danieli} {et~al.}(2020){Danieli}, {van Dokkum}, {Abraham}, {Conroy},
  {Dolphin}, \& {Romanowsky}}]{Danieli2020ApJL}
{Danieli}, S., {van Dokkum}, P., {Abraham}, R., {et~al.} 2020, \apjl, 895, L4,
  \dodoi{10.3847/2041-8213/ab8dc4}

\bibitem[{{Danieli} {et~al.}(2019){Danieli}, {van Dokkum}, {Conroy}, {Abraham},
  \& {Romanowsky}}]{Danieli2019ApJL}
{Danieli}, S., {van Dokkum}, P., {Conroy}, C., {Abraham}, R., \& {Romanowsky},
  A.~J. 2019, \apjl, 874, L12, \dodoi{10.3847/2041-8213/ab0e8c}

\bibitem[{{Doppel} {et~al.}(2021){Doppel}, {Sales}, {Navarro}, {Abadi}, {Peng},
  {Toloba}, \& {Ramos-Almendares}}]{Doppel2021MNRAS}
{Doppel}, J.~E., {Sales}, L.~V., {Navarro}, J.~F., {et~al.} 2021, \mnras, 502,
  1661, \dodoi{10.1093/mnras/staa3915}

\bibitem[{{Emsellem} {et~al.}(2019){Emsellem}, {van der Burg}, {Fensch},
  {Je{\v{r}}{\'a}bkov{\'a}}, {Zanella}, {Agnello}, {Hilker}, {M{\"u}ller},
  {Rejkuba}, {Duc}, {Durrell}, {Habas}, {Lelli}, {Lim}, {Marleau}, {Peng}, \&
  {S{\'a}nchez-Janssen}}]{Emsellem2019A&A}
{Emsellem}, E., {van der Burg}, R. F.~J., {Fensch}, J., {et~al.} 2019, \aap,
  625, A76, \dodoi{10.1051/0004-6361/201834909}

\bibitem[{{Fensch} {et~al.}(2019){Fensch}, {van der Burg},
  {Je{\v{r}}{\'a}bkov{\'a}}, {Emsellem}, {Zanella}, {Agnello}, {Hilker},
  {M{\"u}ller}, {Rejkuba}, {Duc}, {Durrell}, {Habas}, {Lim}, {Marleau}, {Peng},
  \& {S{\'a}nchez Janssen}}]{Fensch2019A&A}
{Fensch}, J., {van der Burg}, R. F.~J., {Je{\v{r}}{\'a}bkov{\'a}}, T., {et~al.}
  2019, \aap, 625, A77, \dodoi{10.1051/0004-6361/201834911}

\bibitem[{{Forbes} {et~al.}(2018){Forbes}, {Read}, {Gieles}, \&
  {Collins}}]{Forbes2018MNRAS}
{Forbes}, D.~A., {Read}, J.~I., {Gieles}, M., \& {Collins}, M. L.~M. 2018,
  \mnras, 481, 5592, \dodoi{10.1093/mnras/sty2584}

\bibitem[{{Gannon} {et~al.}(2023){Gannon}, {Forbes}, {Brodie}, {Romanowsky},
  {Couch}, \& {Ferr{\'e}-Mateu}}]{Gannon2023MNRAS}
{Gannon}, J.~S., {Forbes}, D.~A., {Brodie}, J.~P., {et~al.} 2023, \mnras, 518,
  3653, \dodoi{10.1093/mnras/stac3264}

\bibitem[{{Gray}(1982)}]{Gray1982ApJ}
{Gray}, D.~F. 1982, \apj, 262, 682, \dodoi{10.1086/160461}

\bibitem[{{Jackson} {et~al.}(2021){Jackson}, {Kaviraj}, {Martin}, {Devriendt},
  {Slyz}, {Silk}, {Dubois}, {Yi}, {Pichon}, {Volonteri}, {Choi}, {Kimm},
  {Kraljic}, \& {Peirani}}]{Jackson2021MNRAS}
{Jackson}, R.~A., {Kaviraj}, S., {Martin}, G., {et~al.} 2021, \mnras, 502,
  1785, \dodoi{10.1093/mnras/stab093}

\bibitem[{{Keim} {et~al.}(2022){Keim}, {van Dokkum}, {Danieli}, {Lokhorst},
  {Li}, {Shen}, {Abraham}, {Chen}, {Gilhuly}, {Liu}, {Merritt}, {Miller},
  {Pasha}, \& {Polzin}}]{Keim2022ApJ}
{Keim}, M.~A., {van Dokkum}, P., {Danieli}, S., {et~al.} 2022, \apj, 935, 160,
  \dodoi{10.3847/1538-4357/ac7dab}

\bibitem[{{Laporte} {et~al.}(2019){Laporte}, {Agnello}, \&
  {Navarro}}]{Laporte2019MNRAS}
{Laporte}, C. F.~P., {Agnello}, A., \& {Navarro}, J.~F. 2019, \mnras, 484, 245,
  \dodoi{10.1093/mnras/sty2891}

\bibitem[{{Lewis} {et~al.}(2020){Lewis}, {Brewer}, \& {Wan}}]{Lewis2020MNRAS}
{Lewis}, G.~F., {Brewer}, B.~J., \& {Wan}, Z. 2020, \mnras, 491, L1,
  \dodoi{10.1093/mnrasl/slz157}

\bibitem[{{Macci{\`o}} {et~al.}(2021){Macci{\`o}}, {Prats}, {Dixon}, {Buck},
  {Waterval}, {Arora}, {Courteau}, \& {Kang}}]{Maccio2021MNRAS}
{Macci{\`o}}, A.~V., {Prats}, D.~H., {Dixon}, K.~L., {et~al.} 2021, \mnras,
  501, 693, \dodoi{10.1093/mnras/staa3716}

\bibitem[{{Massarotti} {et~al.}(2008){Massarotti}, {Latham}, {Stefanik}, \&
  {Fogel}}]{Massarotti2008AJ}
{Massarotti}, A., {Latham}, D.~W., {Stefanik}, R.~P., \& {Fogel}, J. 2008, \aj,
  135, 209, \dodoi{10.1088/0004-6256/135/1/209}

\bibitem[{{Montes} {et~al.}(2021){Montes}, {Trujillo}, {Infante-Sainz},
  {Monelli}, \& {Borlaff}}]{Montes2021ApJ}
{Montes}, M., {Trujillo}, I., {Infante-Sainz}, R., {Monelli}, M., \& {Borlaff},
  A.~S. 2021, \apj, 919, 56, \dodoi{10.3847/1538-4357/ac0d55}

\bibitem[{{Moreno} {et~al.}(2022){Moreno}, {Danieli}, {Bullock}, {Feldmann},
  {Hopkins}, {{\c{c}}atmabacak}, {Gurvich}, {Lazar}, {Klein}, {Hummels},
  {Hafen}, {Mercado}, {Yu}, {Jiang}, {Wheeler}, {Wetzel},
  {Angl{\'e}s-Alc{\'a}zar}, {Boylan-Kolchin}, {Quataert},
  {Faucher-Gigu{\`e}re}, \& {Kere{\v{s}}}}]{Moreno2022NatAs}
{Moreno}, J., {Danieli}, S., {Bullock}, J.~S., {et~al.} 2022, Nature Astronomy,
  6, 496, \dodoi{10.1038/s41550-021-01598-4}

\bibitem[{{M{\"u}ller} {et~al.}(2019){M{\"u}ller}, {Famaey}, \&
  {Zhao}}]{Muller2019A&A_mond}
{M{\"u}ller}, O., {Famaey}, B., \& {Zhao}, H. 2019, \aap, 623, A36,
  \dodoi{10.1051/0004-6361/201834914}

\bibitem[{{Nadler} {et~al.}(2020){Nadler}, {Wechsler}, {Bechtol}, {Mao},
  {Green}, {Drlica-Wagner}, {McNanna}, {Mau}, {Pace}, {Simon}, {Kravtsov},
  {Dodelson}, {Li}, {Riley}, {Wang}, {Abbott}, {Aguena}, {Allam}, {Annis},
  {Avila}, {Bernstein}, {Bertin}, {Brooks}, {Burke}, {Rosell}, {Kind},
  {Carretero}, {Costanzi}, {da Costa}, {De Vicente}, {Desai}, {Evrard},
  {Flaugher}, {Fosalba}, {Frieman}, {Garc{\'\i}a-Bellido}, {Gaztanaga},
  {Gerdes}, {Gruen}, {Gschwend}, {Gutierrez}, {Hartley}, {Hinton}, {Honscheid},
  {Krause}, {Kuehn}, {Kuropatkin}, {Lahav}, {Maia}, {Marshall}, {Menanteau},
  {Miquel}, {Palmese}, {Paz-Chinch{\'o}n}, {Plazas}, {Romer}, {Sanchez},
  {Santiago}, {Scarpine}, {Serrano}, {Smith}, {Soares-Santos}, {Suchyta},
  {Tarle}, {Thomas}, {Varga}, {Walker}, \& {DES Collaboration}}]{Nadler2020ApJ}
{Nadler}, E.~O., {Wechsler}, R.~H., {Bechtol}, K., {et~al.} 2020, \apj, 893,
  48, \dodoi{10.3847/1538-4357/ab846a}

\bibitem[{{Nusser}(2020)}]{Nusser2020ApJ}
{Nusser}, A. 2020, \apj, 893, 66, \dodoi{10.3847/1538-4357/ab792c}

\bibitem[{{Ogiya}(2018)}]{Ogiya2018MNRAS}
{Ogiya}, G. 2018, \mnras, 480, L106, \dodoi{10.1093/mnrasl/sly138}

\bibitem[{{Ogiya} {et~al.}(2022){Ogiya}, {van den Bosch}, \&
  {Burkert}}]{Ogiya2022MNRAS}
{Ogiya}, G., {van den Bosch}, F.~C., \& {Burkert}, A. 2022, \mnras, 510, 2724,
  \dodoi{10.1093/mnras/stab3658}

\bibitem[{{Sheminova}(2019)}]{Sheminova2019KPCB}
{Sheminova}, V.~A. 2019, Kinematics and Physics of Celestial Bodies, 35, 129,
  \dodoi{10.3103/S088459131903005X}

\bibitem[{{Shen} {et~al.}(2021{\natexlab{a}}){Shen}, {van Dokkum}, \&
  {Danieli}}]{Shen2021ApJ}
{Shen}, Z., {van Dokkum}, P., \& {Danieli}, S. 2021{\natexlab{a}}, \apj, 909,
  179, \dodoi{10.3847/1538-4357/abdd29}

\bibitem[{{Shen} {et~al.}(2021{\natexlab{b}}){Shen}, {Danieli}, {van Dokkum},
  {Abraham}, {Brodie}, {Conroy}, {Dolphin}, {Romanowsky}, {Kruijssen}, \&
  {Dutta Chowdhury}}]{Shen2021ApJL}
{Shen}, Z., {Danieli}, S., {van Dokkum}, P., {et~al.} 2021{\natexlab{b}},
  \apjl, 914, L12, \dodoi{10.3847/2041-8213/ac0335}

\bibitem[{{Silk}(2019)}]{Silk2019MNRAS}
{Silk}, J. 2019, \mnras, 488, L24, \dodoi{10.1093/mnrasl/slz090}

\bibitem[{{Spitler} \& {Forbes}(2009)}]{Spitler2009MNRAS}
{Spitler}, L.~R., \& {Forbes}, D.~A. 2009, \mnras, 392, L1,
  \dodoi{10.1111/j.1745-3933.2008.00567.x}

\bibitem[{{Toloba} {et~al.}(2023){Toloba}, {Sales}, {Lim}, {Peng},
  {Guhathakurta}, {Roediger}, {Wang}, {Mihos}, {C{\^o}t{\'e}}, {Durrell}, \&
  {Ferrarese}}]{Toloba2023ApJ}
{Toloba}, E., {Sales}, L.~V., {Lim}, S., {et~al.} 2023, \apj, 951, 77,
  \dodoi{10.3847/1538-4357/acd336}

\bibitem[{{van Dokkum} {et~al.}(2019{\natexlab{a}}){van Dokkum}, {Danieli},
  {Abraham}, {Conroy}, \& {Romanowsky}}]{vanDokkum2019ApJL}
{van Dokkum}, P., {Danieli}, S., {Abraham}, R., {Conroy}, C., \& {Romanowsky},
  A.~J. 2019{\natexlab{a}}, \apjl, 874, L5, \dodoi{10.3847/2041-8213/ab0d92}

\bibitem[{{van Dokkum} {et~al.}(2016){van Dokkum}, {Abraham}, {Brodie},
  {Conroy}, {Danieli}, {Merritt}, {Mowla}, {Romanowsky}, \&
  {Zhang}}]{vanDokkum2016ApJL}
{van Dokkum}, P., {Abraham}, R., {Brodie}, J., {et~al.} 2016, \apjl, 828, L6,
  \dodoi{10.3847/2041-8205/828/1/L6}

\bibitem[{{van Dokkum} {et~al.}(2018){van Dokkum}, {Danieli}, {Cohen},
  {Merritt}, {Romanowsky}, {Abraham}, {Brodie}, {Conroy}, {Lokhorst}, {Mowla},
  {O'Sullivan}, \& {Zhang}}]{vanDokkum2018Natur}
{van Dokkum}, P., {Danieli}, S., {Cohen}, Y., {et~al.} 2018, \nat, 555, 629,
  \dodoi{10.1038/nature25767}

\bibitem[{{van Dokkum} {et~al.}(2019{\natexlab{b}}){van Dokkum}, {Wasserman},
  {Danieli}, {Abraham}, {Brodie}, {Conroy}, {Forbes}, {Martin}, {Matuszewski},
  {Romanowsky}, \& {Villaume}}]{vanDokkum2019ApJ}
{van Dokkum}, P., {Wasserman}, A., {Danieli}, S., {et~al.} 2019{\natexlab{b}},
  \apj, 880, 91, \dodoi{10.3847/1538-4357/ab2914}

\bibitem[{{van Dokkum} {et~al.}(2022{\natexlab{a}}){van Dokkum}, {Shen},
  {Romanowsky}, {Abraham}, {Conroy}, {Danieli}, {Chowdhury}, {Keim},
  {Kruijssen}, {Leja}, \& {Trujillo-Gomez}}]{vanDokkum2022ApJL}
{van Dokkum}, P., {Shen}, Z., {Romanowsky}, A.~J., {et~al.} 2022{\natexlab{a}},
  \apjl, 940, L9, \dodoi{10.3847/2041-8213/ac94d6}

\bibitem[{{van Dokkum} {et~al.}(2022{\natexlab{b}}){van Dokkum}, {Shen},
  {Keim}, {Trujillo-Gomez}, {Danieli}, {Dutta Chowdhury}, {Abraham}, {Conroy},
  {Kruijssen}, {Nagai}, \& {Romanowsky}}]{vanDokkum2022Natur}
{van Dokkum}, P., {Shen}, Z., {Keim}, M.~A., {et~al.} 2022{\natexlab{b}}, \nat,
  605, 435, \dodoi{10.1038/s41586-022-04665-6}

\bibitem[{{Wolf} {et~al.}(2010){Wolf}, {Martinez}, {Bullock}, {Kaplinghat},
  {Geha}, {Mu{\~n}oz}, {Simon}, \& {Avedo}}]{Wolf2010MNRAS}
{Wolf}, J., {Martinez}, G.~D., {Bullock}, J.~S., {et~al.} 2010, \mnras, 406,
  1220, \dodoi{10.1111/j.1365-2966.2010.16753.x}

\bibitem[{{Zaritsky} \& {Behroozi}(2023)}]{Zaritsky2023MNRAS}
{Zaritsky}, D., \& {Behroozi}, P. 2023, \mnras, 519, 871,
  \dodoi{10.1093/mnras/stac3610}

\end{thebibliography}
\bibliographystyle{aasjournal}



\end{document}